\documentclass[twocolumn]{aastex631}

\usepackage{orcidlink}
\usepackage{graphics,epsf}
\usepackage[utf8]{inputenc}
\usepackage{amsmath}                
\usepackage{amsfonts}               
\usepackage{amssymb}                
\usepackage{epsfig}                 
\usepackage{graphicx}
\usepackage{float}
\usepackage{color}
\usepackage{multirow}               
\usepackage{hyperref}
\usepackage{xspace}

\hypersetup{
    colorlinks=true,
    linkcolor=red,   
    urlcolor=cyan}

\usepackage[para,online,flushleft]{threeparttable}

\usepackage[colorinlistoftodos]{todonotes}

\newcommand{\s}{{~\rm s}}

\newcommand{\g}{{~\rm g}}
\newcommand{\G}{{~\rm G}}
\newcommand{\K}{{~\rm K}}
\newcommand{\erg}{{~\rm erg}}
\newcommand{\yr}{{~\rm yr}}

\newcommand{\kpc}{{~\rm kpc}}

\newcommand{\keV}{{~\rm keV}}

\begin{document}

\title{Explaining supernova remnant G352.7-0.1 as a peculiar type Ia supernova inside a planetary nebula}


\author{Noam Soker\,\orcidlink{0000-0003-0375-8987}} 
\affiliation{Department of Physics, Technion, Haifa, 3200003, Israel;  soker@physics.technion.ac.il}

\begin{abstract}
I identify a point-symmetric morphology of the supernova remnant (SNR) G352.7-0.1 and propose that the outer axially-symmetric structure is the remnant of a common envelope evolution (CEE) of the progenitor system, while the inner structure is the ejecta of a thermonuclear explosion triggered by the merger of a white dwarf (WD) and the core of an asymptotic giant branch (AGB) star. The main radio structure of SNR G352.7-0.1 forms an outer (large) ellipse. The bright X-ray emitting gas forms a smaller ellipse with a symmetry axis inclined to the symmetry axis of the large radio ellipse. The high abundance of iron and the energy of its X-ray lines suggest a type Ia supernova (SN Ia). The massive swept-up gas suggests a relatively massive progenitor system. I propose a scenario with progenitors of initial masses of  $M_{\rm ZAMS,1} \simeq 5-7 M_\odot$ and  $M_{\rm ZAMS,2} \simeq 4-5 M_\odot$. At a later phase, the WD remnant of the primary star and the AGB secondary star experience a CEE that ejects the circumstellar material that swept up more ISM to form the large elliptical radio structure. An explosion during the merger of the WD with the core of the AGB star triggered a super-Chandrasekhar thermonuclear explosion that formed the inner structure that is bright in X-ray.  A tertiary star in the system caused the misalignment of the two symmetry axes. This study adds to the wide variety of evolutionary routes within the scenarios of normal and peculiar SNe Ia.
\end{abstract}


\section{Introduction} 
\label{sec:intro}

The morphology of a supernova remnant (SNR) can reveal its progenitor and the explosion mechanism of the parent supernova. This holds (for a recent review see \citealt{Soker2024Rev}) for both core-collapse supernovae (CCSNe) and type Ia supernovae (SNe Ia). One prominent morphological feature is point symmetry, where two or more pairs of twin structural components are on opposite sides of the center of the nebula and the symmetry axes of different pairs are inclined to each other.   
Axially symmetric structures in SNRs might result from the circumstellar material (CSM) that the progenitor of the supernova blows before the explosion and/or from the explosion itself.

In the case of SNe Ia, which are exploding white dwarfs (WDs), the CSM is ionized by a WD (or even two) before the explosion and hence it is a planetary nebula. An SN inside a planetary nebula is termed SNIP (e.g., \citealt{TsebrenkoSoker2015SNIP}). Many planetary nebulae are known to possess axially-symmetric or point-symmetric morphologies  (e.g.,  \citealt{Balick1987, Chuetal1987, Schwarzetal1992, CorradiSchwarz1995, Manchadoetal1996, SahaiTrauger1998, Parkeretal2016, Parker2022}, for some catalogs). These point-symmetric morphologies of planetary nebulae and post-asymptotic giant branch (AGB) nebulae are not always perfect (e.g., \citealt{Sahaietal2000, Sahaietal2011, Hrivnaketal2001}). Namely, the two opposite sides might be unequal in their distance from the center of the nebula, their brightness, their exact shapes, and/or their sizes. The point-symmetry of the type Ia SNR G1.9+0.3 suggests that it is due to the point-symmetric morphology of the planetary nebula into which a spherical SN Ia explosion took place \citep{Soker2023G1903}.

There are several SN Ia scenarios (for reviews from the last decade that cover many aspects of SNe Ia and hundreds of references therein see \citealt{Maozetal2014, MaedaTerada2016, Hoeflich2017, LivioMazzali2018, Soker2018Rev, Soker2019Rev, WangB2018,  Jhaetal2019NatAs, RuizLapuente2019, Ruiter2020, Aleoetal2023, Liuetal2023Rev, Vinkoetal2023, Soker2024Rev}). 
In \cite{Soker2019Rev} and \cite{Soker2024Rev} I classified these scenarios according to whether their explosion ejecta is spherical or not. I argued there that for a spherical explosion ejecta, the explosion is of a lonely WD. 

A \textit{lonely SN Ia scenario} is one where at the explosion time itself there is only a WD and without any close companion. The WD itself is a descendant of a close binary interaction, i.e., binary merger. Two scenarios belong to the lonely WD group, the core degenerate (CD) scenario where a WD merges with the core of an AGB star during a common envelope evolution (CEE) and the explosion occurs with a merger to explosion delay (MED) time, and the double degenerate (DD) with MED scenario where two WDs merge but the explosion occurs after a MED time (the DD-MED scenario; for the DD-MED scenario see also \citealt{Neopaneetal2022}). 
The DD scenario, with a prompt explosion in a WD binary system, leads to a non-spherical explosion (e.g.,  \citealt{Pakmoretal2012, Pakmoretal2013, Pakmoretal2022, Peretsetal2019, Zenatietal2023}). 
The WD-WD collision (WWC) scenario also leads to non-spherical explosions (e.g., \citealt{Kushniretal2013, Glanzetal2024}). An axially-asymmetric explosion might result from the DD scenario (without a MED time; e.g., \citealt{Peretsetal2019}), or possibly from a rapidly rotating lonely WD. 

In this study, I examine the morphology of SNR G352.7-0.1, and identify a point-symmetric morphology with two symmetry axes (section \ref{sec:PointSymmetry}). 
\cite{ToledoRoyetal2014} conduct three-dimensional hydrodynamical simulations of a supernova explosion inside a cloud that is bounded by an interstellar medium (ISM) to reproduce the morphology of SNR G352.7-0.1. \cite{Zhangetal2023} consider their model unlikely because it is not compatible with the molecular clouds' properties with which SNR G352.7-0.1 interacts. Also, \cite{ToledoRoyetal2014}, who did not consider jets,  did not produce any symmetry axis. The topics of the present study are the two symmetry axes of SNR G352.7-0.1 and their implications on the progenitor. 

There is a debate on whether G352.7-0.1 is a CCSN or an SN Ia (section \ref{sec:Debate}). I accept the view that it is an SN Ia and use the identified point symmetry to suggest a peculiar SN Ia scenario for SNR G352.7-0.1 (section \ref{sec:Scenario}) where one axis is due to the pre-explosion CSM and the other to the non-spherical explosion. I summarize this study and place the results on a broader scope in section \ref{sec:Summary}.     

\section{The debate over the nature of SNR G352.7-0.1} 
\label{sec:Debate}

\begin{figure*}
\begin{center}
\includegraphics[trim=0.0cm 6.1cm 0.0cm 1.0cm ,clip, scale=0.80]{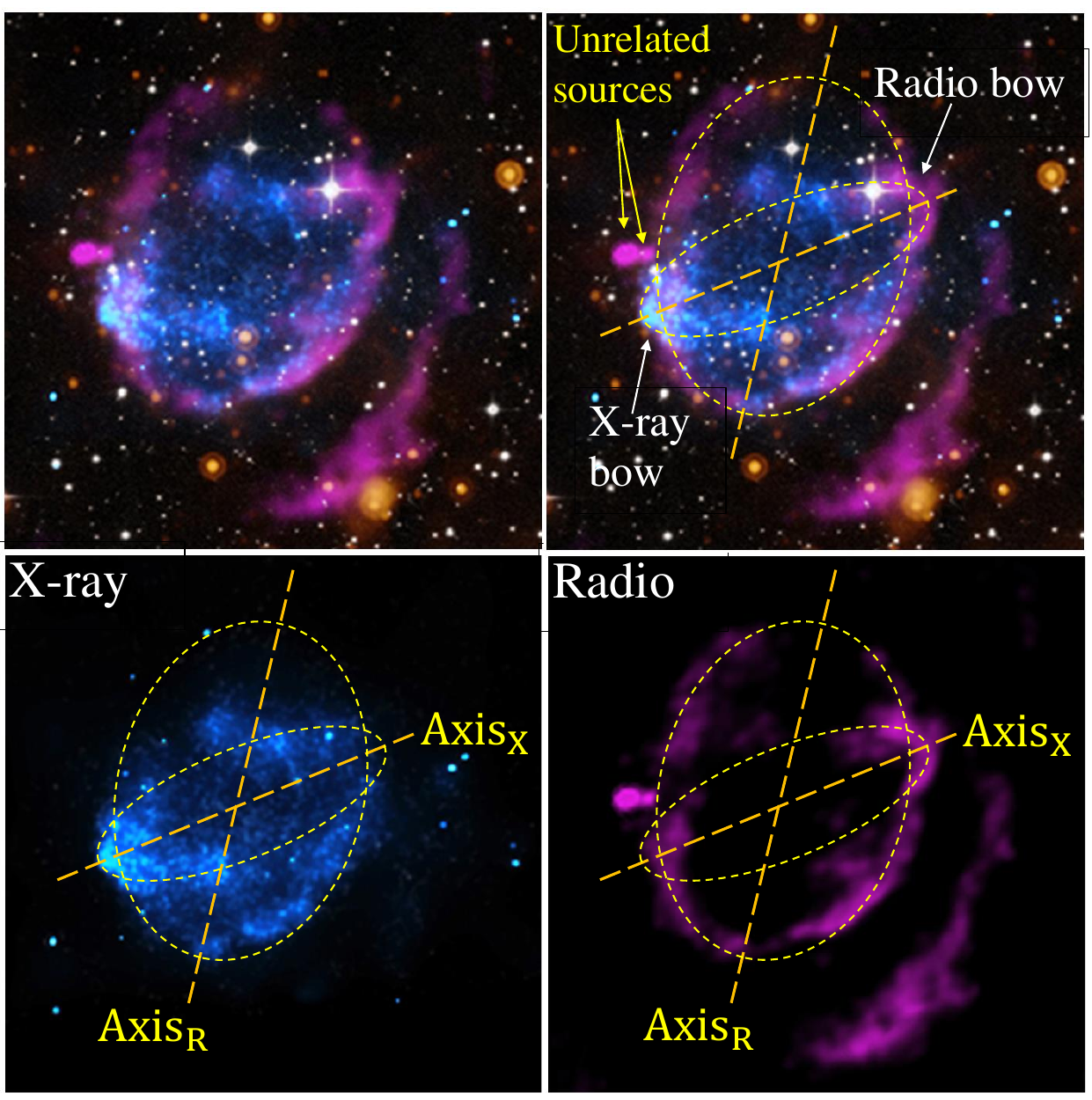}
\end{center}
\caption{
Images from the Chandra X-Ray Observatory homepage (https://chandra.harvard.edu/photo/2014/g352/). In all panels X-ray emission is shown in blue, optical emission is shown in grayscale, infrared emission is shown in orange, and radio emission is shown in pink. The lower-left panel is a Chandra image in the $0.5–8.0 \keV$ band from \cite{Pannutietal2014}. I added the two ellipses on three panels and the identification of two symmetry axes, one by the radio image ($\rm{Axis}_{\rm R}$) and one by the X-ray image (${\rm Axis}_{\rm X}$).  
Credit for the images: 	X-ray: NASA/CXC/Morehead State Univ/T.Pannuti et al.; Optical: DSS; Infrared: NASA/JPL-Caltech; Radio: NRAO/VLA/Argentinian Institute of Radioastronomy/G.Dubner). 
}
\label{Fig:Figure1} 
\end{figure*}

In Figure \ref{Fig:Figure1} I present the X-ray (lower left panel), radio (lower right panel), and a composite image (upper panels) of SNR G352.7-0.1. I return to this figure and its analysis in section \ref{sec:PointSymmetry}.

The age of SNR G352.7-0.1 is about several thousand years (e.g., \citealt{Fujishigeetal2023}). The SNR is well resolved in the X-ray (e.g., \citealt{Kinugasaetal1998, Giacanietal2009}) and the radio (e.g., \citealt{Caswelletal1983, Dubneretal1993, Giacanietal2009}) bands. The X-ray and radio morphologies are not identical, and hence this is classified as a mixed-morphology SNR (e.g., \citealt{Vink2012} for the mixed morphology classification).  
While some earlier studies argue that SNR G352.7-0.1 is the remnant of a CCSN, some recent studies attribute SNR G352.7-0.1 to an SN Ia. 

\cite{Giacanietal2009} suggest that SNR G352.7-0.1 is a descendant of a CCSN. They based this claim on the asymmetrical morphology and the massive swept-up gas (see also \citealt{Pannutietal2014}). \cite{Yamaguchietal2014} classified SNRs according to the ionization state of iron and found SNR G352.7-0.1 to be an SNR Ia. \cite{SezerGok2014} and \cite{Fujishigeetal2023} analyze Suzaku X-ray observations of SNR G352.7-0.1. Based on the high iron abundance they derive, \cite{SezerGok2014} argue that SNR G352.7-0.1 resulted from an SN Ia. \cite{Fujishigeetal2023} obtain plasma temperatures, ionization parameters, and abundances, that are different from those that \cite{SezerGok2014} obtained. Nonetheless, they also claim that SNR G352.7-0.1 is a descendant of an SN Ia. Another support for the SN Ia classification is the non-detection of a central compact object;  \cite{Pannutietal2014} could not find a central compact object from \textit{Chandra} observations. 

\cite{Zhangetal2023} study the interaction of SNR G352.7-0.1 with molecular clouds. Based on the general classification by \cite{Manchester1987} of SNR radio morphologies, they suggest that SNR G352.7-0.1 morphology is due mainly to the progenitor wind. 
They estimate the duration of the wind as $\approx 3 \times 10^5 \yr$ and the total energy of the progenitor wind to be $\approx 3 \times 10^{48} \erg$, and note that these are compatible with winds of O stars, in the case of a CCSN origin of SNR G352.7-0.1. For an SN Ia explosion, they conclude that the wind of an asymptotic giant branch (AGB) star cannot account for this energy, and they suggest a wind from the WD progenitor that accreted mass at a high rate from an AGB star and blew a strong wind. In that case, one expects an AGB star inside the SNR. \cite{Zhangetal2023} do not take a clear side on whether SNR G352.7-0.1  was a CCSN or an SN Ia. 

I will attribute the high energy of the CSM of SNR G352.7-0.1 to the ejection of a common envelope from a relatively massive system, a WD of $M_{\rm WD} \simeq 1 M_\odot$ spiraling-in inside the envelope of an AGB star of $M_{\rm AGB} \simeq 5-7 M_\odot$. 
The zero-age main sequence (ZAMS) progenitor of SNR G352.7-0.1 was a binary system of relatively massive stars (with a low mass tertiary star; section \ref{sec:Scenario}). Namely, the primary star was on the upper mass range that forms a CO WD, $M_{\rm ZAMS,1} \simeq 5-7 M_\odot$, and the secondary star was less massive, $M_{\rm ZAMS,2} \simeq 4-5 M_\odot$, but could have accreted mass from the primary before leaving the main sequence and hence evolve faster than had it not accreted mass.  
Such a secondary star evolves to the AGB in $\tau_{\rm ev} \simeq 10^8 \yr$. This relatively short evolution time accounts for the presence of molecular clouds that this SNR interacts with, despite possibly being a descendant of an SN Ia.

I am postponing more details of the proposed peculiar supernova scenarios to section \ref{sec:Scenario}. In the next section, I reveal the two symmetry axes that I use to construct the possible peculiar scenarios for SNR G352.7-0.1.   

\section{Revealing a point-symmetry in a composite X-ray/radio images} 
\label{sec:PointSymmetry}

Because the morphologies in radio and X-ray observations differ, SNR G352.7-0.1 is classified as a mixed-morphology SNR. Based on the recognition that many SNRs have two or more morphological features with symmetry axes that are inclined to each other, i.e., point-symmetric SNRs (section \ref{sec:intro}), I examine the possibility that the mixed morphology of SNR G352.7-0.1 is a point symmetric morphology. 

\begin{figure}
\begin{center}
\includegraphics[trim=0.6cm 14.1cm 0.0cm 1.0cm ,clip, scale=0.64]{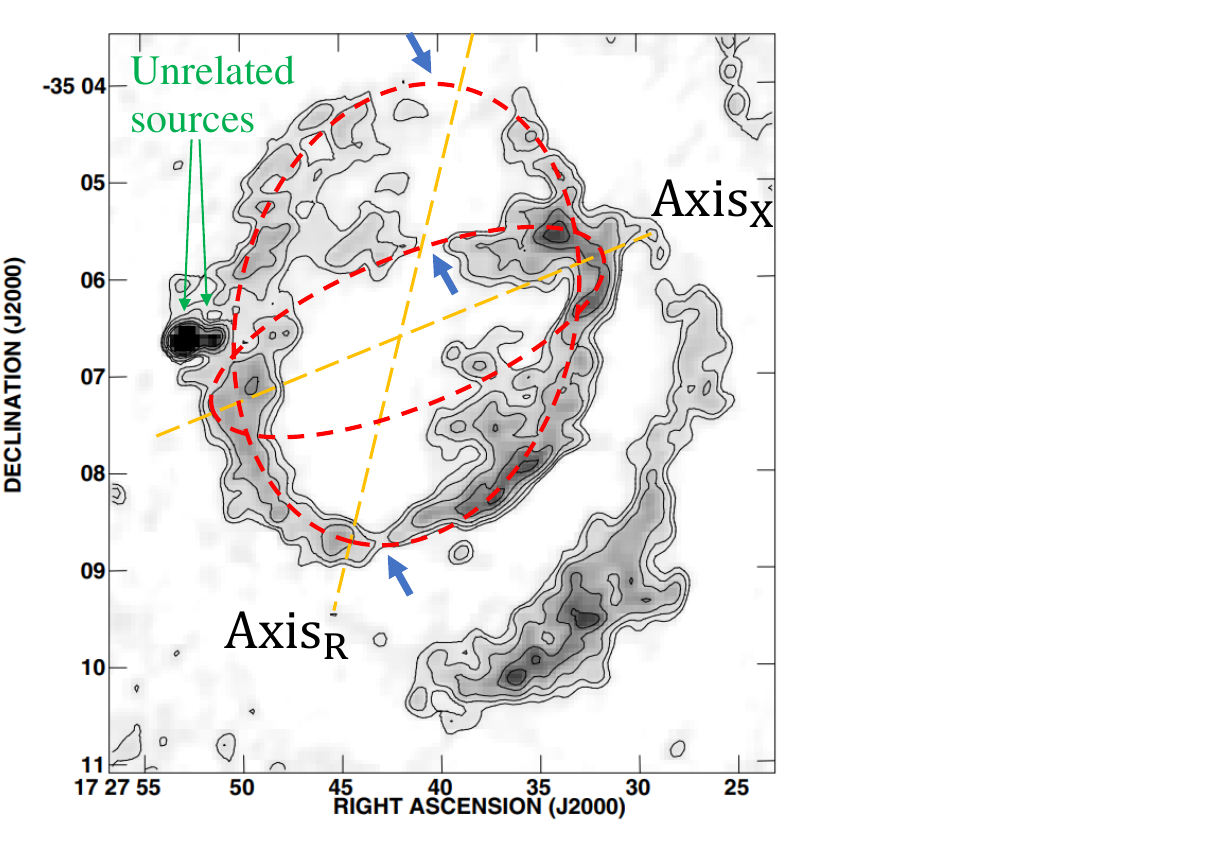}
\end{center}
\caption{
Radio images from \cite{Giacanietal2009}. I added the two symmetry axes and the ellipses that are identical to those in Figure \ref{Fig:Figure1}. The three blue arrows point at minimum radio intensity in the corresponding filaments. The two unrelated point-like sources are WBH2005 352.775-0.153 and WBH2005 352.772-0.149 \citep{Whiteetal2005}. }
\label{Fig:Figure2} 
\end{figure}

\begin{figure*}
\begin{center}
\includegraphics[trim=0.6cm 22.1cm 0.0cm 0.0cm ,clip, scale=0.80]{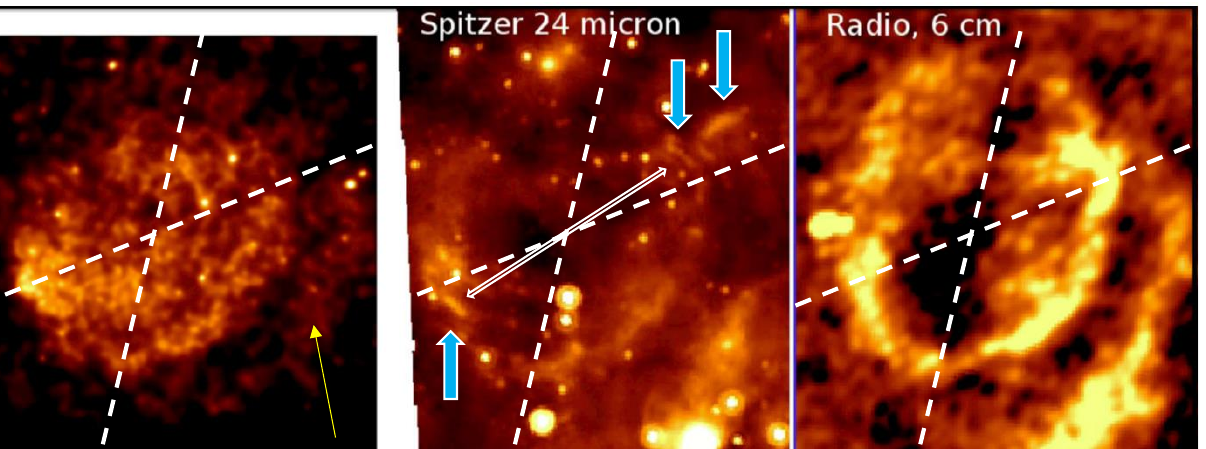}
\end{center}
\caption{ Figure 9 from \cite{Pannutietal2014} with my notations of the two symmetry axes, in dashed-white lines, that I identify in Figures \ref{Fig:Figure1} and \ref{Fig:Figure2}. From left to right: X-ray (Chandra in $0.5–8.0 \keV$), infrared (Spitzer in 24 micron), and radio (VLA 6 cm). The axis of the larger ellipse goes through the minimum in radio intensity zones (two black regions in the right panel). I also added three thick arrows pointing at opposite (relative to the center) infrared clumps. I drew a double-headed arrow pointing at the opposite clumps, which practically coincides with the axis of the smaller ellipse. 
The yellow arrow on the left panel points at the outer X-ray arc.}
\label{Fig:Figure3} 
\end{figure*}

In the upper panels of Figure \ref{Fig:Figure1} I present a composite image from the Chandra X-Ray Observatory homepage of X-ray, radio, infrared (IR), and visible light. On the upper-right panel and the lower panels, I added my identification of two symmetry axes (dashed straight lines). The large ellipse marks the approximate bright radio ellipse, best seen in the radio image on the lower-right panel. I fit the ellipses and their axes by eyes, using the radio image in Figure \ref{Fig:Figure1} and that from \cite{Giacanietal2009} that I present in Figure \ref{Fig:Figure2}. In Figure \ref{Fig:Figure2} the blue short arrows point at three minima along filaments in the radio intensity map. The symmetry axis of the large ellipse very closely coincides with the line that goes through the three minima. This strengthens the identification of this symmetry axis. The same axis through the radio-minima, the ${\rm Axis}_{\rm R}$ one, is seen in the right panel of Figure \ref{Fig:Figure3} as it crosses the two elongated minima in radio intensity (dark zones).  

The identification of the second symmetry axis is more delicate. To identify the symmetry axis of the smaller ellipse and to draw the smaller ellipse, I use both the X-ray and radio morphologies. I identify a bow-shaped radio filament and a less clear bow-shaped X-ray zone (or bright X-ray clump), to be two opposite structural features that were shaped together. I term them the radio bow and X-ray bow as I mark on the upper right panel of Figure \ref{Fig:Figure1}. I take the center of these two bows to define the axis of a smaller ellipse and draw a corresponding ellipse according to the X-ray map. 

\cite{Giacanietal2009} suggest that SNR G352.7-0.1 has a barrel-shaped morphology. The axis of the barrel includes the radio shell (large ellipse in Figures \ref{Fig:Figure1} and \ref{Fig:Figure2}), and the outer radio arc on the southwest (the incomplete southwest radio arc on the lower right of radio images), and does not coincide with either of the two symmetry axes that I identify. The projection on the plane of the sky of a circular ring on a tilted barrel is an ellipse. The two opposite tips of the longer axis of the projected ellipse (upper and lower segments of the large ellipse in Figure \ref{Fig:Figure2}) should be brighter than the rest of the ellipse (because a longer line of sight through the ring relative to the two perpendicular segments on the ring). However, the opposite holds for the larger radio ellipse of Figure \ref{Fig:Figure2}. I therefore prefer to consider the outer radio arc as a structure further away from the center, rather than a symmetric part to the southwest segment of the radio ellipse.  

As evident from the literature, SNR G352.7-0.1 is puzzling in being non-spherical and having a massive shell, properties typical to CCSNe, but with high iron abundance and no central compact object, properties typical to SNRs Ia. The identification of two symmetry axes in an SNR that is already puzzling makes this a very rare case.
I could not find a similar case. 
The closest SNR in morphology is the SNR Ia DEM L71 in the Large Magellanic Cloud (for recent images see \citealt{Lietal2021, AlanBilir2022}). DEM L71 has an elliptical structure with ears and a clear axis. However, it is less puzzling and I cannot identify a clear second declined axis. Still, since it is classified as a SNIP \citep{Soker2022CEED}, it might well be that the interaction that shaped the PN into which the progenitor of SNR DEM L71 exploded was shaped by triple star interaction. But not necessarily so. 

I turn to propose a scenario that accounts for the properties of SNR G352.7-0.1.  

\section{Possible scenarios for SNR G352.7-0.1} 
\label{sec:Scenario}

As I discuss in section \ref{sec:Debate} the iron properties of SNR G352.7-0.1, mass and line energy, suggest an SN Ia, while the highly non-spherical morphology, the massive swept-up mass, and the molecular clouds in the ISM suggest a massive stellar system progenitor. 
The location of SNR G352.7-0.1 towards the galactic center, $(l,b)=(352^\circ.7579, -0^\circ.1236)$ and an estimated distance of $6.8-8.4 \kpc$ \citep{Fujishigeetal2023} and up to $\approx 10.5 \kpc$ \citep{Zhangetal2023}, is compatible with possible recent star formation and a massive stellar system progenitor, i.e., on the upper mass zone for formation of CO WDs.  

To these properties, I add my identification of two symmetry axes, as I mark on the Figures and discuss in section \ref{sec:PointSymmetry}.  
\cite{Pannutietal2014} estimate the mass of the X-ray emitting gas to be $2.6 M_\odot$ and that of the swept-up ISM mass to be $45 M_\odot$. The high ISM mass suggests that the ISM substantially influenced the SNR morphology. Nonetheless, I attribute the two symmetry axes that I identify to shaping by the progenitor and the explosion and not to the ISM. 

To incorporate the above properties I suggest that SNR G352.7-0.1 is either a peculiar SN Ia or a peculiar SN II (one that is not a CCSN), in both cases the explosion is thermonuclear in a super-Chandrasekhar merger of two WDs. 

The evolutionary phases and basic ingredients of the possible scenarios, as well as their relations to the observed properties of SNR G352.7-0.1, are as follows. 
\begin{enumerate}
\item \textit{The Progenitor triple system.}
The fundamental components of the progenitor are two stars with initial masses of $M_{\rm ZAMS,1} \simeq 5-7 M_\odot$ and  $M_{\rm ZAMS,2} \simeq 4-5 M_\odot$. In addition, there was a tertiary star around the primary star with an orbital plane that was inclined to that of the main (outer) binary system (see point 5 below). The secondary stellar evolution time to its AGB phase is $\approx 10^8 \yr$. During this period the system did not move to a large distance from its formation location and the ISM that supplied gas to star formation did not have time to fully disperse. This accounts for the dense ISM and the molecular clouds in the vicinity of SNR G352.7-0.1. 
\item {\textit{Formation of the WD.}}
The primary star formed a massive CO WD of $M_{\rm WD,1} \simeq 1-1.1 M_\odot$ (by initial-final mass relation, e.g., \citealt{Marigo2022, Cunninghametal2024}). The secondary star avoided CEE before it left the main sequence. However, the secondary star could have accreted mass to become a star of $M_2 \simeq 6-7 M_\odot$. The tertiary lower-mass main sequence star did enter a CEE and ended closer to the newly born WD. The inner binary system, composed of the WD and the tertiary star, could have been a cataclysmic variable. The mass that the primary star blew during its evolution was already mixed with the ISM before the secondary star left the main sequence.    
\item \textit{Common envelope evolution.} As the secondary became an AGB star it engulfed the WD and its main sequence low-mass star companion, and the system entered a CEE. The triple system ejected most or all of the envelope of the secondary star.
\item \textit{The outcome of CEE and the explosion.} After the formation of the CEE, there are three possible scenarios. 
\begin{enumerate}
    \item 
\textbf{Peculiar SN Ia via the DD scenario.} The system manages to eject the entire envelope and the WD and the core end at a very close orbit that brings them to merge within $t_{\rm CEED}  \simeq {\rm few} \times 10^5 \yr$. This relatively short CEE to explosion delay (CEED) time (see \citealt{Soker2022CEED} for the introduction of the CEED time) implies that the explosion ejecta collides with the CSM that was formed in the CEE.
The merger is driven by emitting gravitational waves, by triple-star interaction, or both. Until the explosion, the central core ionizes the CSM which is the ejected common envelope, and the system is a planetary nebula. As the WD and the core merge, they explode. This is the DD scenario (without a MED) that results in a peculiar SN Ia because the combined mass of the core and the WD is $\simeq 1.6-2.2 M_\odot$, i.e., super-Chandrasekhar. This is an SN Ia inside a planetary nebula (SNIP). In the merging of the core with the WD, one of them is tidally destroyed and forms a disk around the other. This disk launches jets that shape the explosion ejecta that is observed in X-ray, i.e., the smaller ellipse in the figures. The axis of the explosion ejecta is inclined to that of the CSM because of the triple-star interaction (see point 5 below). 
\item
\textbf{Peculiar SN Ia via the CD scenario.} The WD and the core merge during or at the end of the CEE, and the merger remnant explodes after envelope ejection but within a CEED time of  $t_{\rm CEED} \simeq {\rm few} \times 10^5 \yr$. This is again an SNIP, but a peculiar SN Ia via the CD scenario. In this case, the explosion is expected to be close to spherical (unless there are two opposite bullets along the rotation axis of the merger remnant). Therefore, this scenario does not account for the elliptically-shaped X-ray ejecta and its symmetry axis. This scenario is not likely for SNR G352.7-0.1. 
\item 
\textbf{Peculiar SN II via the CD scenario.} In this scenario the WD and the core merge at the end of the CEE and explode before the entire envelope is ejected or shortly after the ejection of the common envelope but before it becomes transparent, a scenario that \cite{LivioRiess2003} studied. Because of the massive hydrogen in the CSM, this might be classified as a peculiar type II supernova. The explosion during the merger process ensures a non-spherical explosion ejecta, as in route (a) above.   
\end{enumerate} 
\item \textit{Accounting for inclined symmetry axes.} The two symmetry axes that I identify are inclined to each other. Above I attributed the material that formed the radio ellipse to the former common envelope that was ejected during the CEE. I attribute the X-ray emitting gas to the explosion ejecta. I suggest that the presence of a tertiary star with an inclined orbit to that of the two main stars accounts for the inclination. Most likely, a lower mass star, $M_3 \simeq 0.3-1 M_\odot$, orbited the primary star. As the primary formed an AGB star, the tertiary might have spiraled-in inside a CEE to a close orbit. The system of the WD and the tertiary might have been a catalytic variable. The tertiary star induced the merger or influenced its plane, such that the WD-core merger plane was inclined to the plane of the primary-secondary system. This deserves further study. I note here that about half of such massive progenitors are likely to be in triple-star systems (e.g., \citealt{MoeDiStefano2017}). 
\item \textit{The fate of the tertiary star.} The tertiary star might end in one of a few ways (e.g., \citealt{SabachSoker2015, GlanzPerets2021}). It can be ejected from the common envelope and become bound or unbound from the binary system. In any case, after the explosion, it will be a lonely star. In principle, it should still be inside the SNR now. More likely, because the tertiary star influenced the merger process, it survived till the final stages of the CEE when it was tidally destroyed by either the core or the WD. In this case, it does not exist anymore. 
\end{enumerate} 

I end this section by speculating on the nature of the incomplete southwestern radio arc (outer radio arc). The lower-left panel of Figure \ref{Fig:Figure1} and the left panel of Figure \ref{Fig:Figure3} show that there is an X-ray emitting gas just inner to the outer radio arc which is disconnected from the inner main X-ray structure. This I term the `outer X-ray arc'; the yellow arrow on the left panel of Figure \ref{Fig:Figure3} points at this arc. I speculate that the shell that makes the main radio ellipse is not complete, and that hot ejecta leaks out and interacts with the outer CSM. The outer CSM is an outer shell, formed by the triple-star interaction during the formation process of the planetary nebula. This interaction forms the outer radio arc. This speculation can be tested with hydrodynamical simulations.   

\section{Summary} 
\label{sec:Summary}

I proposed that the explosion that formed SNR G352.7-0.1 was a super-Chandrasekhar thermonuclear explosion during the merger of two degenerate stars. 
The explosion occurred during the merger process of two WDs in a DD scenario (route-a in point 4 of section \ref{sec:Scenario}) that resulted in a peculiar super-Chandrasekhar SN Ia or during the merger of the core and the WD at the end of a CEE in what could have been observed as a peculiar type II SN, although it is a super-Chandrasekhar thermonuclear explosion as well (route-c there). The super-Chandrasekhar thermonuclear explosion accounts for the relatively massive ejecta as revealed by the X-ray emitting gas ($\simeq 2.6 M_\odot$, \citealt{Pannutietal2014}), the high iron abundance,  and the energy of iron X-ray lines (section \ref{sec:Debate}).  

The relatively massive progenitors, $M_{\rm ZAMS,1} \simeq 5-7 M_\odot$ and  $M_{\rm ZAMS,2} \simeq 4-5 M_\odot$, account for the SNR being in a region of dense ISM and the presence of molecular clouds (section \ref{sec:Debate}). The pre-explosion CEE of the WD remnant of the primary star inside the AGB envelope of the secondary star formed a CSM that is the larger ellipse observed in radio (Figures \ref{Fig:Figure1} and \ref{Fig:Figure2}). 
The explosion ejecta is the material observed in the X-ray, which defines the smaller ellipse and its symmetry axis. 

To account for the misalignment of the two symmetry axes that I identify in this study, I suggest the presence of a tertiary star (point 5 in section \ref{sec:Scenario}). 

In \cite{Soker2023G1903} I attributed the point-symmetric morphology of SNR G1.9+0.3 solely to the planetary nebula into which a spherical normal SN Ia explosion took place. In the case of SNR G352.7-0.1 that I study here, on the other hand, the point symmetry is attributed both to a planetary nebula progenitor and a non-spherical peculiar SN Ia explosion. 

Although not common, the proposed scenario is not extreme in its ingredients and properties. Firstly, in the progenitor binary system mass range, $M_{\rm ZAMS,1} \simeq 5-7 M_\odot$, triple star systems are common, i.e., about half of the binary systems are triple and higher order stellar systems (e.g., \citealt{MoeDiStefano2017}).  Secondly, normal SN Ia explosions within several hundred thousand years after the CEE, $t_{\rm CEED} \la {\rm several} \times 10^5 \yr$ are common. The gas that was blown during the CEE forms a planetary nebula as it is ionized by the WD remnant. The explosion inside a planetary nebula is an SNIP. In \cite{Soker2022CEED} I estimated that the fraction of SNIPs among normal SNe Ia is $f_{\rm SNIP} \simeq 0.5$. This study hints that peculiar SNIPs are also common. 

The present study adds to the rich variety of evolutionary routes within the different scenarios of normal and peculiar SNe Ia. 

\section*{Acknowledgements}

I thank an anonymous referee for useful comments. This research was supported by a grant from the Israel Science Foundation (769/20).


\end{document}